# Grain incompatibility determines the local structure of amorphous grain boundary complexions

Pulkit Garg [a], Timothy J. Rupert [a,b,*]

[a] Department of Materials Science and Engineering, University of California, Irvine, CA 92697, USA

[b] Department of Mechanical and Aerospace Engineering, University of California, Irvine, CA 92697, USA

* To whom correspondence should be addressed: trupert@uci.edu

Amorphous grain boundary complexions lack long-range crystalline order but are not featureless, as distinct gradients in structural short-range order have been reported through their thickness. In this work, we test the hypothesis that the distribution of short-range order is determined by the confining crystals using atomistic simulations of both Cu-Zr bicrystals and a random polycrystal. Voronoi polyhedra with structures similar to that of perfect face-centered cubic serve as signatures of high structural order and are only found at the amorphous-crystalline transition regions. The density of the ordered structural motifs within a specific amorphous-crystalline transition region is found to not be directly determined by the orientation and symmetry of the grain which touches it, but rather by the incompatibility between the two confining grains. Ordered polyhedra density is found to be inversely related to grain incompatibility, meaning that large incompatibilities between the confining crystals lead to less order in the amorphous-crystalline transition region. The finding that the entire grain-film-grain system must be considered to understand local structure unequivocally demonstrates that amorphous complexions are not simply a collection of independent phases which happen to nucleate at a grain boundary. Rather, an amorphous grain boundary complexion is a single entity that finds a local equilibrium configuration.

# 1. Introduction

In polycrystalline metals, bulk properties are determined by the atomic details associated with the crystallography of the grains and the structure of grain boundaries [1]. For example, grain boundaries act as obstacles to dislocation motion [2, 3] and thus play an important role in determining mechanical properties such as strength, ductility, creep, and fatigue resistance [4, 5]. However, grain boundaries are high energy regions as compared to the crystalline bulk, which promotes grain growth that can diminish the properties of polycrystalline metals [6, 7]. Tuning the interfacial network to reduce the density of boundaries with detrimental properties and increase the number of "special" boundaries with beneficial properties is one of the methods to restrict grain growth and/or reduce negative behaviors associated with grain boundaries, such as corrosion [8-10]. The low energy of special boundaries is often attributed to the shared periodicity of the two grain lattices at the interface and can be classified using analytical tools such as the coincident site lattice (CSL) model [11, 12]. While there is some understanding of how boundary properties are associated with macroscopic crystallographic parameters such as grain misorientation and grain boundary plane normal, the dependence of boundary properties on the details of interfacial structure are not yet fully understood.

Solute segregation is another commonly used method to decorate grain boundaries and manipulate their behavior, with increased stability against grain growth [13-15] and improved strength [16-18] being well-known positive outcomes. Interfacial segregation is a complex process as it depends on grain boundary structure, and each of the crystallographic parameters can potentially be important [19-21]. Solute segregation can modify interfacial structure [22-24] and also lead to structural transitions of grain boundaries between different interfacial phases, also known as *grain boundary complexions* [25-28]. Grain boundary complexions represent an interfacial structure that is in thermodynamic equilibrium with its abutting grains and typically these features have a stable, finite thickness [29, 30]. Complexions can be structurally ordered or disordered depending on their local atomic arrangement [27], and chemical order and disorder can also exist in multicomponent systems [31-33]. Ordered complexions such as bilayers have been reported to be brittle in normally ductile metals like Al, Cu, and Ni, leading to premature intergranular fracture [34, 35]. Failure of ordered complexions to absorb dislocations during plastic deformation can also promote crack formation and intergranular fracture in nanocrystalline metals [36-38]. In contrast, structurally disordered or amorphous complexions can efficiently absorb dislocations during plastic deformation and enhance mechanical damage resistance by delaying crack nucleation and suppressing crack propagation in nanocrystalline metals [39, 40]. Grain boundary complexions with amorphous structure have also been observed to act as high-capacity dislocation sinks and increase the tensile ductility of nanocrystalline layered films [41, 42]. Amorphous complexions have been shown to improve thermal stability [43] while simultaneously increasing the strength and ductility of nanocrystalline Cu-Zr alloy as compared to pure nanocrystalline Cu [44, 45], addressing the two minor weaknesses which are typical of nanocrystalline materials. Thicker amorphous complexions lead to further improvements in the ductility of nanocrystalline alloys, as they can accommodate more incoming dislocations before cracking [46, 47]. In general, the results above show that amorphous grain boundary complexions offer the possibility to engineer polycrystalline alloys with improved strength and ductility, yet further study is needed to understand how the nuances of amorphous complexion structure can be tuned and manipulated to optimize this type of behavior.

As bulk amorphous metals have been studied in great detail, these materials serve as a useful starting point for a discussion of structure in disordered complexions, which can be seen as interfacial analogs of bulk metallic glasses. Several models such as the dense random packing model [48, 49], stereo-chemically defined model [50], efficient local atomic packing model [51], and efficient cluster-packing model [52] have been proposed to understand the disordered structure of metallic glasses. The most common approach to analyze the local structure of metallic glasses is to characterize the atomic arrangement within the first coordination shell, with structural short-range order (SRO) defined as the reoccurrence of the same type of packing motifs [53, 54]. There is significant variety in terms of SRO type as well as the packing densities of these structural motifs in metallic glasses [55, 56]. However, SRO has indeed been connected to many important properties of bulk metallic glasses, providing a specific target feature [57]. In Cu-Zr metallic glasses, icosahedral clusters consisting of 12 first coordination shell atoms with five-fold symmetry [58, 59] have been directly connected to various properties such as configurational potential energy, specific heat, and local shear modulus [60, 61]. Furthermore, icosohedral SRO is closely related to the glass-forming ability [62, 63], transport [64], and mechanical properties [65, 66] of Cu-Zr metallic glasses. For example, an increase in the density of icosahedral SRO can provide more resistance to plastic flow, leading to increased strength and ductility of bulk amorphous metals [67]. During deformation, icosahedral clusters break down into distorted and unstable clusters which can combine to form shear bands that lead to the fracture of metallic glasses [68, 69]. While geometrically-favored icosahedral SRO plays a key role in the deformation response of bulk amorphous metals, the geometrically unfavored clusters also contribute to shear banding by forming disordered soft regions prone to instability [70, 71]. Many unusual properties of bulk amorphous metals have also been attributed to medium-range order originating from the arrangement and connection of SRO units in glassy materials [72, 73]. The network of SRO units has been observed to evolve during cooling [74], forming stronger connections between SRO units with decreasing cooling rates [53, 75].

Similar SRO motifs have been observed in grain boundary complexions between different types of abutting crystals [76]. Amorphous complexions have been proposed by some authors to be a different class of interfaces with only two crystallographic parameters needed to describe their shear/slip transfer properties, as compared to grain boundaries which require five crystallographic parameters [41, 77, 78], although such a claim has not yet been confirmed. In addition to grain boundary character, temperature can also play a critical role in interfacial segregation, the transition between complexion types [24, 79], the thickening of interfacial complexions [80, 81], and the formation of wetting films [25, 27]. However, essential features of amorphous complexion structure such as SRO patterning and its dependence on the confining crystals have not been clarified to date. Pan and Rupert [82] provided preliminary work on this topic by simulating grain boundary complexions in Cu-Zr symmetric tilt bicrystal samples, with these authors describing the disordered complexion as two structurally distinct regions (the amorphous interior and the amorphous-crystalline transition regions (ACTRs)), each with different types of packing motifs dominating the SRO in each region and separated by a transition region in some cases. Interfacial segregation has been shown to strongly depend on boundary structure [35, 83, 84] with segregation induced structures depending on the crystallography of neighboring grains [85], motivating a more thorough study of amorphous complexion structure and its dependence on the confining grains. Understanding the impact of initial grain structure on the resultant amorphous complexion state will enable the manipulation of polycrystalline microstructures to develop nanocrystalline alloys with optimized properties.

In this study, atomistic simulations are used to investigate the role of neighboring grains on the local atomic structure of amorphous complexions in a Cu-Zr alloy, with a particular eye for understanding how the incompatibility between abutting crystals determines complexion structure. Hybrid Monte Carlo (MC)/molecular dynamics (MD) simulations are first used to create equilibrated Cu-Zr bicrystal samples with different neighboring grains and nanoscale complexions at three different temperatures, with a polycrystalline model providing additional data points. The complexions are analyzed using a Voronoi tessellation method [86-88] which provides detailed information about SRO distribution and has been successfully used in previous studies to describe bulk amorphous phases [58, 59, 67]. Unique Voronoi polyhedra corresponding to the amorphous interior and the ACTRs that separate the amorphous interior from the neighboring crystalline grains are identified. Interestingly, the density of ordered SRO motifs at either ACTR is found to be asymmetric in the asymmetric tilt and twist grain boundaries equilibrated at 600 K and 1000 K, yet this asymmetry is not present in the wetting film samples equilibrated at 1600 K (i.e., above the alloy melting temperature). In contrast, symmetric grain boundaries always produce two ACTRs with matching ordered SRO densities. This asymmetry/symmetry in ACTR SRO highlights a connection between the local structural order of grain boundary complexions and the neighboring grains. By considering different boundary crystallographies, a relationship between the ACTR SRO and the incompatibility between the neighboring grains is clearly demonstrated, showing definitively that amorphous grain boundary complexions are one system that finds a local equilibrium configuration.

## 2. Computational Methods

Complexion formation was simulated with hybrid atomistic MC/MD simulations using the Large-scale Atomic/Molecular Massively Parallel Simulator (LAMMPS) code [89], with a 1 fs integration time step for all MD simulations. An improved version [90] of the previously available [91] embedded-atom method potential for Cu-Zr is used in this work. The newly calibrated potential prevents nucleation of artificial Laves phases and therefore describes an amorphous Cu-Zr structure more accurately. All atomic configurations were visualized using the open-source visualization tool OVITO [92].

First, a reference pure Cu bicrystal sample was created to represent an asymmetric high angle grain boundary, with the X-axis aligned along the [1 1 0] direction of Grain 1 and the [1 1 $\bar{1}$] direction of Grain 2 (named HAGB-1 and shown in Fig. 1). The two grains in HAGB-1 are oriented perpendicular to each other to provide a high degree of incompatibility. To further test our hypothesis that the confining crystals affect the grain boundary structure, nine additional types of grain boundaries with different combinations of neighboring grains were examined, with details presented in Table 1. The grain boundaries are referred to using $\Sigma$ values, representing the coincidence sites between the two neighboring grains, in conjunction with the CSL model [12, 93]. Symmetric boundaries with the same orientation of the two confining crystals include low ($\Sigma$61a), medium ($\Sigma$5), and high ($\Sigma$25) angle symmetric tilt grain boundaries, and low ($\Sigma$61-Tw) and medium angle ($\Sigma$29-Tw) symmetric twist grain boundaries. On the other hand, another high angle grain boundary with grains perpendicular to each other (HAGB-2) and three asymmetric tilt boundaries with low ($\Sigma$5), medium ($\Sigma$5), and high ($\Sigma$9) angles are selected to illustrate the changes in grain boundary structure as the incompatibility between the grains changes. Finally, a polycrystalline sample with a random grain boundary network was simulated to provide a more random collection of grain boundaries. A pure Cu polycrystalline sample was constructed using the Voronoi tessellation method and equilibrated using conjugate gradient minimization and Nose-Hoover thermo/barostat at 900 K. The equilibrated pure Cu polycrystal was then doped with 2 at.% Zr at 900 K using the hybrid MC/MD method to allow solute segregation and subsequent transition of grain boundaries to complexions, with additional details found in Garg et al. [79]. The equilibrated polycrystal contains a network of asymmetric grain boundaries with different types of randomly oriented neighboring grains to further investigate the structure of amorphous complexions.

Table 1. The CSL Σ number, grain boundary normal vectors, and thickness of different types of grain boundaries that are generated from the initial grain boundary structure with different combinations of neighboring grains. Σ61a, Σ5, and Σ25 are low, medium, and high angle symmetric tilt grain boundaries, respectively, and Σ61 Tw and Σ29 Tw are two different symmetric twist grain boundaries. HAGB-1 and HAGB-2 are two different high angle grain boundaries where grain 1 and grain 2 are oriented at an angle of 90º. The other three asymmetric tilt grain boundaries are Σ5, Σ5, and Σ9 with low, medium, and high misorientation angles, respectively.

| Grain boundary structure | | Grain boundary thickness (nm) | | |
|---|---|---|---|---|
| CSL Σ number | Grain boundary normal $(hkl)_1/(hkl)_2$ | Low-T complexion | High-T complexion | Wetting Film |
| Σ61a | (0 1 11)/(0 1 11) | 6.6 | 7.0 | 7.6 |
| Σ5 | (0 1 3)/(0 1 3) | 7.7 | 8.3 | 9.2 |
| Σ25 | (0 4 3)/(0 4 3) | 7.0 | 7.2 | 8.0 |
| Σ61 Tw | (11 1 0)/(11 $\bar{1}$ 0) | 6.4 | 7.6 | 8.3 |
| Σ29 Tw | (5 2 0)/(5 $\bar{2}$ 0) | 6.0 | 7.8 | 8.8 |
| **HAGB-1** | (1 1 0)/(1 $\bar{1}$ 1) | 7.7 | 7.9 | 9.5 |
| **HAGB-2** | (1 1 0)/(0 0 1) | 5.7 | 8.9 | 10.4 |
| Σ5 | (7 4 0)/(8 1 0) | 7.0 | 7.2 | 7.5 |
| Σ5 | (6 7 0)/(9 2 0) | 6.8 | 6.9 | 7.6 |
| Σ9 | (1 1 0)/(7 7 8) | 5.9 | 7.1 | 7.9 |

The bicrystal samples were first equilibrated at 600 K and 1000 K using a Nose-Hoover thermo/barostat at zero pressure and then doped with Zr solutes using the hybrid MC/MD method. MC steps were performed in a variance-constrained semi-grand canonical ensemble [94] after every 100 MD steps while maintaining the global composition fixed to 2 at.% Zr by adjusting the chemical potential difference during the simulations [79, 95]. An equilibrated state was achieved when the system's potential energy variation over the last 4000 MC steps is less than 0.001 eV/step and no considerable changes were observed in the composition or structure of the interfacial regions. Fig. 1(a) shows a Cu-Zr bicrystal sample containing two amorphous complexions nucleated at 600 K. Atoms are colored according to their local crystal structure as identified by adaptive common neighbor analysis [96, 97], with face-centered cubic (FCC) atoms colored green and all other types of atoms colored white. The simulation cell is approximately 106 nm long (X-direction), 6 nm tall (Y-direction), and 5 nm thick (Z-direction), containing ~300,000 atoms, with periodic boundary conditions in all three directions. The grain boundary in the bicrystal sample that is doped and equilibrated at 600 K is referred to as a "low-T complexion," while the boundary doped and equilibrated at 1000 K is referred to as a "high-T complexion" in the remaining text. At temperatures above the bulk melting point, grain boundary regions can further transition to wetting films which are bulk liquid phases that happen to reside at the interfacial region [25, 27].

Wetting films were created here by melting the atoms inside the grain boundary regions at 1600 K (above the pure Cu melting temperature of 1353 K for this potential) for 200 ps while the rest of the atoms (i.e., the crystalline grain interiors) are held fixed at 1000 K. The melted atoms are then quenched from 1600 K to 1000 K over 600 ps to retain the high temperature wetting film structure, followed by the removal of the fixed atom constraint on grain interior atoms to smoothly equilibrate the system further.

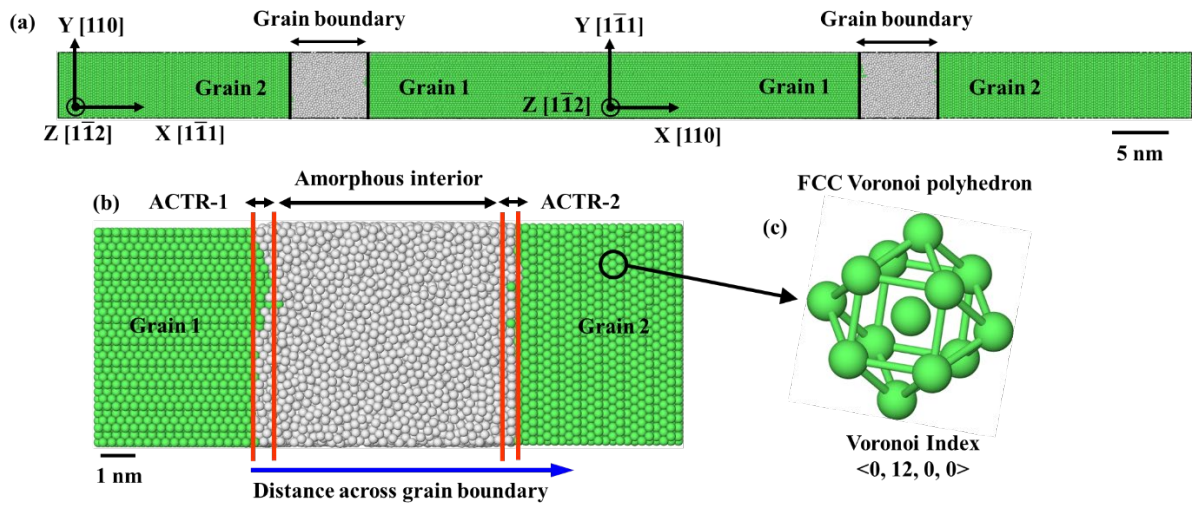

Fig. 1. (a) A Cu-Zr bicrystal sample containing two grain boundaries, prepared from HAGB-1, with the neighboring grains having different orientations. (b) Zoomed-in image of one grain boundary shows an amorphous interior that is confined by two ACTRs. The ACTRs are defined as ACTR-1 and ACTR-2 depending on the neighboring Grain 1 and Grain 2, respectively. (c) Voronoi polyhedron of an atom in the grain interior with perfect FCC crystal structure and Voronoi index <0, 12, 0, 0>. The central atom is surrounded by 12 nearest neighbor atoms forming a polyhedron with 12 faces, each of which has 4 edges.

The structure of a metallic glass is typically sensitive to cooling rate as the distribution and connections of SRO units have been observed to evolve during this process [74]. Significantly decreasing the cooling rate can lead to the formation of a denser network of icosahedral structural motifs [53, 75] which can play a role in the mechanical properties of metallic glasses [98-100]. One could hypothesize that the cooling rates may affect the structure of grain boundary complexions, therefore the effect of this processing variable is also explored here. The low-T complexion, high-T complexion, and wetting film samples were quenched to 10 K at three different cooling rates of $10^{12}$ K/s, $10^{11}$ K/s, and $10^{10}$ K/s. In addition, rapidly quenched samples were also prepared using conjugate gradient energy minimization to remove thermal noise and mimic an extremely fast quenching procedure. Although the cooling rates studied here are significantly higher than typical experimental studies, these rates are typical of MD simulations and consistent with previous studies focused on metallic glass structure [53, 75, 101, 102]. The cooling simulations were performed in the isothermal-isobaric ensemble where the volume of the simulation cell is allowed to change to yield zero pressure in all directions. To damp out any thermal fluctuation from the terminating thermostat and to obtain a stress-free configuration, the systems were relaxed using a Nose–Hoover thermo/barostat at 10 K for 100 ps under zero pressure as a final step. A conjugate gradient minimization was then performed for each sample to remove any remaining thermal noise while preserving the targeted defect structures. The resulting thickness of low-T and high-T complexions ranges from ~6-8 nm while the wetting films have a thickness of ~7.5-9 nm, as shown in Table 1. To provide a useful reference state, bulk metallic glass samples were also produced through the same hybrid MC/MD technique in a periodic simulation box having dimensions of ~13 × 6 × 5 $nm^3$, with quenching treatments performed at the same cooling rates used for the bicrystal samples. The Zr compositions in the bulk amorphous phases were fixed at 17 at.%, 13 at.%, and 11 at.%, to match the grain boundary compositions of the bicrystal samples corresponding to the low-T complexion, high-T complexion, and wetting films, respectively.

The local structure of each atom is characterized using the Voronoi tessellation method [86, 87]. This method divides space into Voronoi polyhedra associated with each atom which are built by constructing bisecting planes along the lines joining the central atom and all of its neighboring atoms. Each Voronoi polyhedron is described using the index notation <$n_3, n_4, n_5, n_6$>, where $n_i$ stands for the number of Voronoi polyhedron faces with $i$ edges and the sum of $n_i$ is the coordination number of the central atom. The density of specific Voronoi polyhedra types can then be a metric used to describe the preferred SRO in the material and has been successfully used in previous studies to understand the structure of amorphous materials [60-63, 65, 66]. Since Cu is the primary elemental species in the alloys studied here, we restrict our discussion to the structural SRO motifs observed around Cu atoms. Structural SRO is based on the local geometric packing of atoms, rather than chemical patterning and ordering which determines chemical SRO.

Local lattice distortions were measured using the elastic strain calculation method in OVITO [103], which analyzes atomic strains in crystalline systems by outputting an elastic strain tensor for each crystalline atom that is measured with respect to the perfect FCC crystal structure. To quantify the overall distortion from the full strain tensor, the von Mises shear strain ($\varepsilon_{VM}$) was calculated for each atom following

$$\varepsilon_{VM} = \left[ \varepsilon_{xy}^2 + \varepsilon_{xz}^2 + \varepsilon_{yz}^2 + 1/6\left((\varepsilon_{xx} - \varepsilon_{yy})^2 + (\varepsilon_{xx} - \varepsilon_{zz})^2 + (\varepsilon_{yy} - \varepsilon_{zz})^2\right)\right]^{1/2} \quad (1)$$

where $\varepsilon_{xx}$, $\varepsilon_{yy}$, $\varepsilon_{zz}$, $\varepsilon_{xy}$, $\varepsilon_{xz}$, and $\varepsilon_{yz}$ are elastic strain components in a symmetric elastic strain tensor [104].

## 3. Results and Discussion

### 3.1 SRO gradients within a general HAGB

To begin to probe the structural order in interfacial regions, we first identify all the different types of polyhedra that are present in the grain boundary regions. Fig. 1(b) shows a grain boundary in the HAGB-1 bicrystal sample containing an amorphous complexion interior (marked by dashed lines) that is bounded by ACTRs. The two ACTRs are defined as ACTR-1 and ACTR-2 because Grain 1 and Grain 2 are the nearest crystal, respectively. The minimum edge length of the Voronoi polyhedra is set to be 0.6 Å to ensure that the polyhedra associated with grain interior atoms are identified as <0, 12, 0, 0>, the building block for perfect FCC crystals. A <0, 12, 0, 0> Voronoi polyhedron consists of a central atom that is surrounded by 12 nearest neighbor atoms forming a polyhedron where each face has 4 edges, as shown in Fig. 1(c).

The internal structure of an amorphous complexion is primarily determined by the most frequent polyhedra in the grain boundary regions and therefore, we focus the discussion and analysis on the most abundant SRO types. Fig. 2(a) shows the atoms in a bulk metallic glass sample with the two most frequent Voronoi polyhedra. The amorphous structure of the metallic glass is dominated by atoms with icosahedral structure or <0, 0, 12, 0> polyhedra, where each of the 12 faces has 5 edges, and near-icosahedral motifs or <0, 2, 10, 0> polyhedra, where 10 faces have 5 edges and the remaining 2 faces have 4 edges [105]. These two motifs will be referred to as "disordered polyhedra" for the remainder of the paper and are shown in Fig. 2(b). They will be used to measure and track the variations in local structure of amorphous regions. Near-icosahedral polyhedra are slightly distorted as compared to perfect icosahedral clusters, but prior work has shown that both SRO types play a similar role in the glass-forming ability, structure, and properties of metallic glasses [106-108]. For example, the disordered structure of bulk amorphous Cu-Zr glasses has been described in terms of the icosahedral and near-icosahedral polyhedra density [109]. Fig. 2(c) shows the atoms with the most common Voronoi polyhedra in an amorphous complexion. The disordered polyhedra <0,0,12,0> and <0,2,10,0> are the two most common motifs in the interior of the complexion, indicating that the SRO in this region is similar to the bulk amorphous phase. In contrast, the SRO in the ACTRs is dominated by <0, 10, 2, 0> and <0, 8, 4, 0> polyhedra. For both <0, 10, 2, 0> and <0, 8, 4, 0> Voronoi polyhedra, each of the central atoms is surrounded by 12 nearest atoms such that the majority of the faces of the polyhedra have 4 edges, as shown in Fig. 2(d). The two dominant SRO types in the ACTRs can be thought of as distorted versions of the perfect FCC polyhedron [110] and will therefore be referred to as "ordered polyhedra" for the remainder of the study to distinguish them from the earlier discussion.

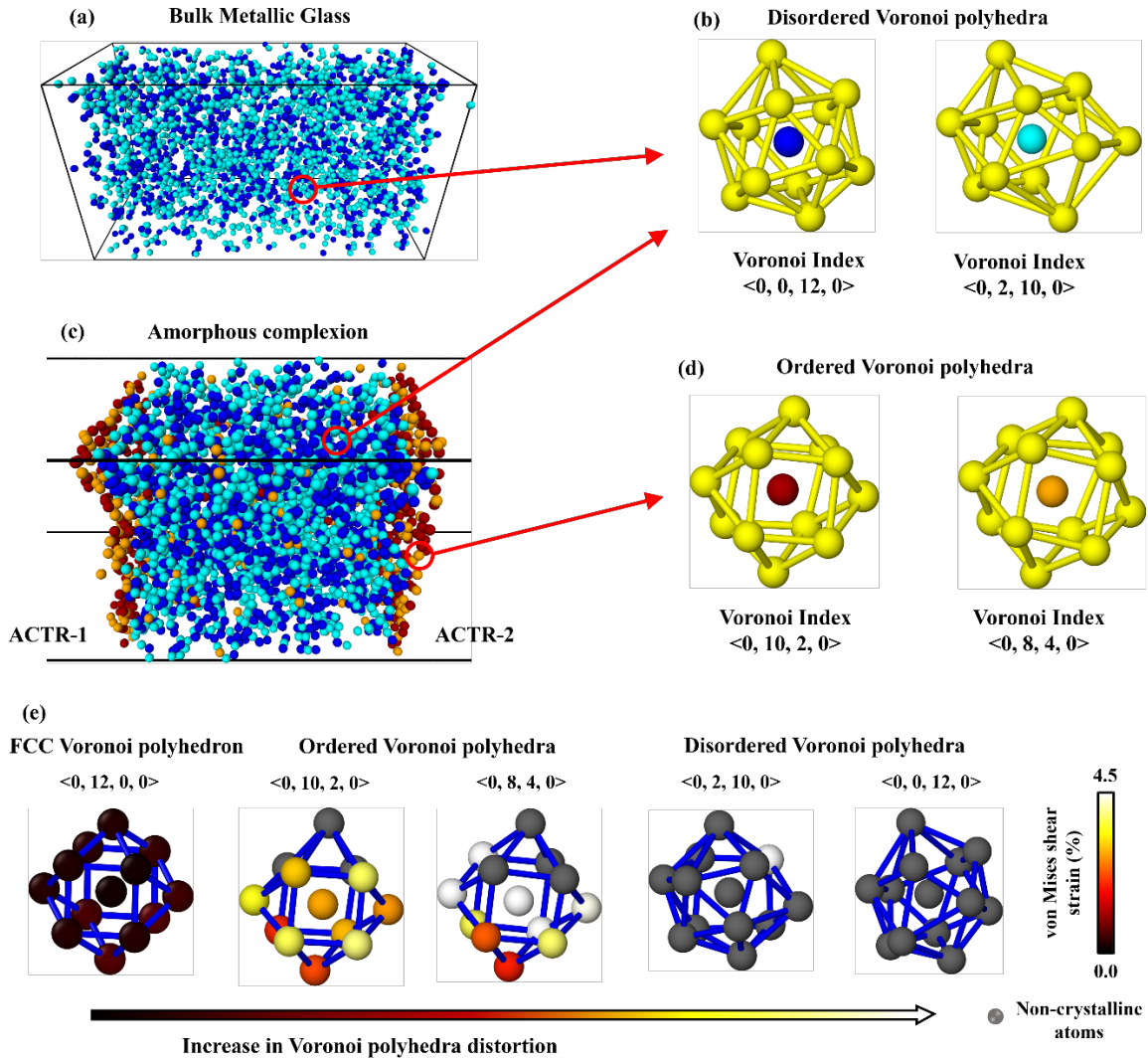

**Fig. 2.** (a) Local structure of bulk metallic glass sample containing atoms with the two most frequent Voronoi polyhedra. (b) Disordered polyhedra <0, 0, 12, 0> and <0, 2, 10, 0> dominate the amorphous structure of bulk metallic glass. (c) Local structure of amorphous complexion, with the four most frequent Voronoi polyhedra in the complexion interior and ACTRs shown. (d) The SRO in ACTRs is dominated by ordered polyhedra <0, 10, 2, 0> and <0, 8, 4, 0> whereas the disordered polyhedra are the most common in the complexion interior. (e) An increase in the local distortional strain is observed within the Voronoi polyhedra as one goes from perfect FCC structure to ordered motifs to disordered Voronoi polyhedra. Thus, ACTRs are more structurally ordered as compared to the complexion interior.

The different SRO polyhedra can conceptually be thought of as existing on a continuum from ordered to disordered, with the former associated with the perfect, crystalline face-centered cubic lattice and the latter associated with a fully glassy phase. Fig. 2(e) shows an increase in the Voronoi polyhedra distortion as one goes from perfect FCC structure to ordered polyhedra to disordered polyhedra. Most of the atoms in the ordered motifs have modest elastic distortions, although a few distorted atoms cannot be mapped back to the perfect FCC structure and are therefore labeled as non-crystalline. In contrast, the disordered motifs primarily consist of fully distorted atoms, and those atoms for which atomic shear strain can be measured exhibit very high values. These observations show that the ACTRs are more structurally ordered as compared to the complexion interior.

Armed with a qualitative view of the SRO distribution in the complexions, we next quantify the density of ordered and disordered SRO motifs as one moves through the complexion in the grain boundary normal direction. Fig. 3 shows the spatial distribution of disordered and ordered SRO types for low-T complexions, high-T complexions, and liquid films quenched at different rates. The ACTR edge can be defined at the location where the ordered polyhedra become more common than the disordered polyhedra, resulting in a thickness of ~0.6 nm for all of the grain boundaries studied in this work. The fraction of Cu atoms with the most frequent polyhedra is calculated by binning the atoms along the X-direction and calculating the ratio of Cu atoms with a given SRO type to the total number of Cu atoms in each bin. A bin size of 0.4 nm, smaller than ACTR thickness to accurately capture SRO gradients, which moves 0.2 nm at every step in the grain boundary starting from one edge of ACTR-1 to the other edge of ACTR-2 was used. The central regions of the plots in Fig. 3 correspond to the amorphous interior with disordered type SRO whereas the shaded regions on the left and right are the ACTRs. For reference, the densities of disordered SRO in the representative bulk amorphous phases of a given composition are shown as black dashed lines with double dots. The distributions of disordered SRO in the interior and ordered SRO at the ACTRs overlap one another for all four cooling rates. An asymmetry of the ordered polyhedra density is observed with ACTR-2 on the right having a higher density of ordered SRO for the low-T and high-T complexion samples shown in Figs. 3(a) and (b), motivating further investigation of the effect of the confining crystals on each side.

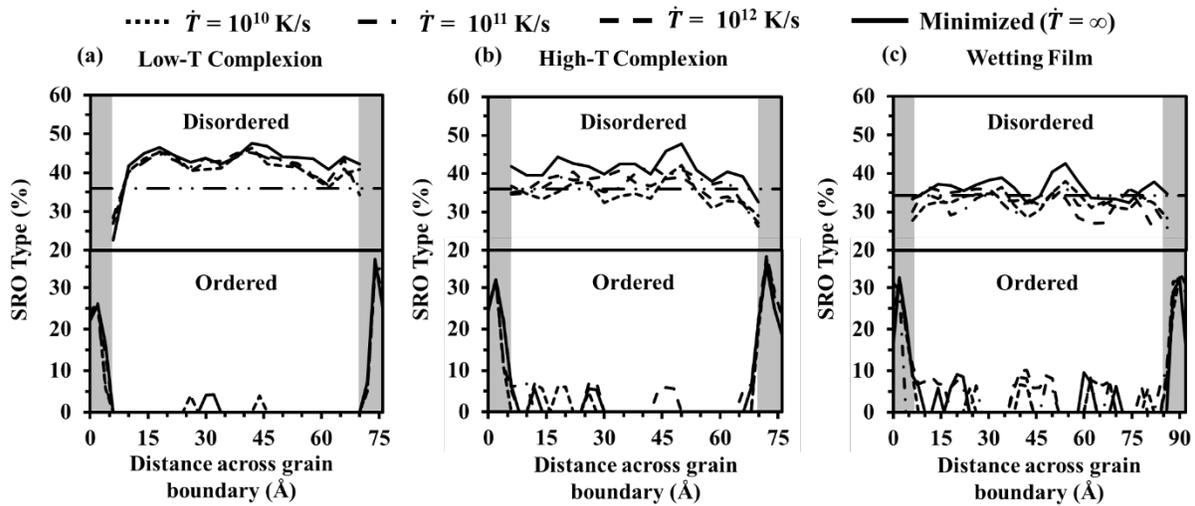

Fig. 3. Spatial variation of SRO types across (a) low-T complexion, (b) high-T complexion, and (c) wetting film models quenched at different cooling rates. The central region of the plots corresponds to the amorphous interior, where disordered polyhedra are the most common. The shaded regions on the left and right are the ACTRs where ordered polyhedra are the most frequent. The black dashed lines with double dots correspond to the spatial distribution of disordered SRO polyhedra in bulk metallic glass samples with Zr concentrations matching the grain boundary regions.

With distinct structural signatures that delineate the ACTR from the complexion interior, it is important to also investigate if there are chemical gradients with the complexions. Fig. 4 shows the distribution of Zr content in the bicrystal simulations, with data from the associated bulk metallic glasses samples represented by dashed lines with double dots. The grain boundary Zr concentration is nearly uniform in the amorphous interior. However, the Zr concentration decreases significantly in the ACTRs. Thus, the increase in ordered SRO polyhedra in the ACTRs is accompanied by a reduction in the Zr dopant concentration. Similar to the observations of SRO type above, spatial trends in the grain boundary Zr concentration are also not affected by the cooling rates used in this study.

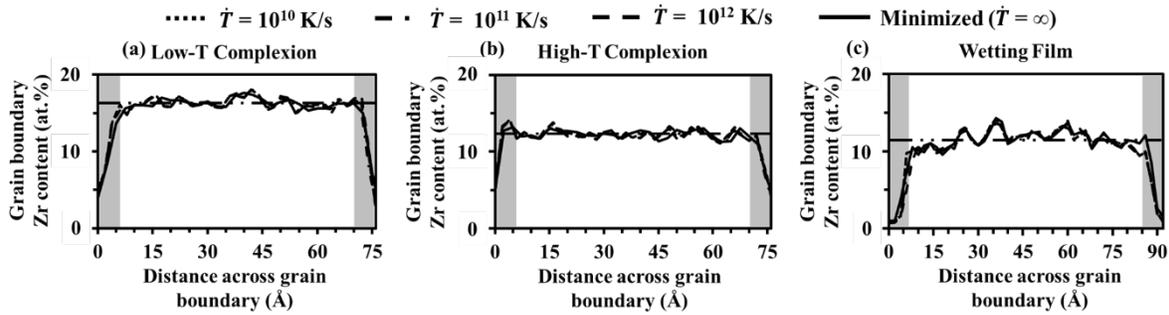

Fig. 4. Grain boundary Zr content across (a) low-T complexion, (b) high-T complexion, and (c) wetting film models quenched at different cooling rates. The shaded regions on the left and right correspond to the ACTR locations and the black dashed lines with double dots correspond to the Zr content in the corresponding bulk metallic glass samples.

To enable a big picture comparison of the ACTR regions, Fig. 5 presents the ordered SRO density, defined as the number of ordered polyhedra centered at ACTR atoms normalized by the volume of the ACTR region, in the various HAGB-1 samples. The ordered SRO density between the two ACTRs is asymmetric with ACTR-1 having a lower ordered SRO density as compared to ACTR-2 in both the low-T and high-T complexion samples. Furthermore, the asymmetry (or gap between the two values for a given set of conditions) is roughly twice as large in the low-T complexion (difference of ~3 atoms/nm$^3$) than in the high-T complexion (difference of ~1.5 atoms/nm$^3$). In contrast, structural asymmetry is not present in the wetting films, as the ordered

SRO density in ACTR-1 and ACTR-2 is identical. Thus, the complexions demonstrate a structural asymmetry that is not present when a true liquid film is simply sandwiched between two grains, signaling that the equilibration between the complete grain-film-grain system is critical to this behavior. The temperature at which grain boundaries are equilibrated also alters the SRO, as the structural asymmetry of the ACTRs is larger in the low-T complexions for the HAGB-1 interface. Again, the ordered SRO density is found to be independent of the cooling rate, at least for the relatively high rates used in this study. Consequently, in the remainder of the paper, we present only the minimized bicrystal samples as these simulations are more computationally efficient and allow us to probe a larger number of grain boundary configurations.

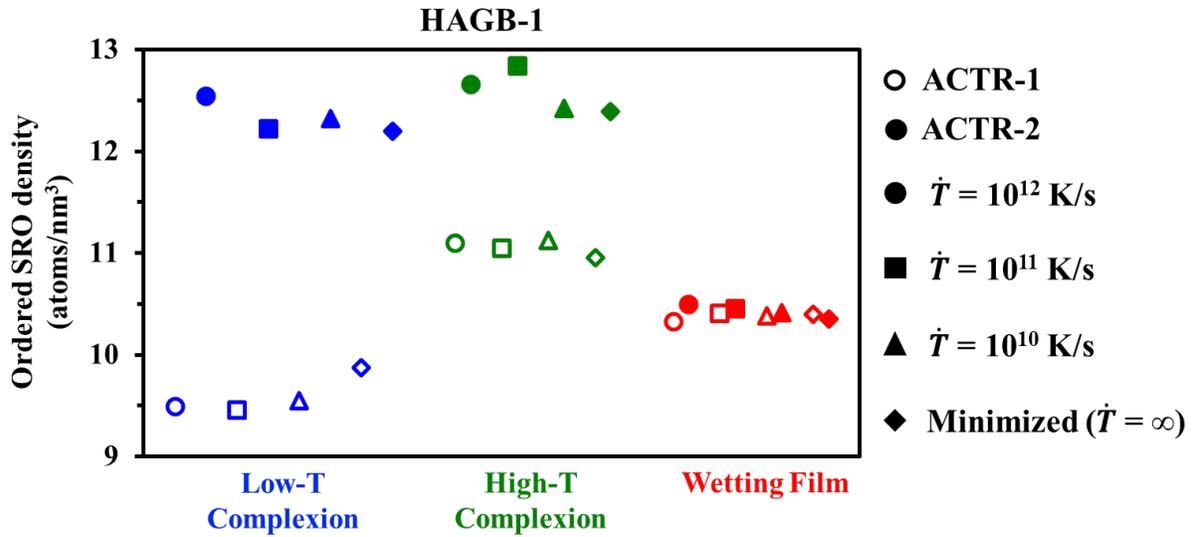

**Fig. 5. Ordered SRO density in the two ACTRs of HAGB-1 with low-T complexion, high-T complexion, and liquid film models quenched at different cooling rates. The ordered SRO density in ACTRs is asymmetric for the low-T and high-T complexions, whereas this asymmetry is not found in the wetting film samples.**

**3.2 Effect of boundary character on SRO distribution**

Having shown that there is an asymmetry in the ordered SRO density at the two ACTRs, we next move to understand the effect of varying the impinging crystals on this behavior. Fig. 6(a) and 6(b) present the ordered SRO density in the ACTRs of low-T complexion, high-T complexion, and wetting film models for symmetric and asymmetric grain boundaries, respectively. For each of the grain boundary configurations, five thermodynamically equivalent configurations (differing by subtle thermal vibrations) are studied to allow for increased statistics. The measured fluctuations in the ordered SRO density are extremely small and represented by error bars in Fig. 6 that are close to the size of the data points. The ordered SRO density in the two ACTRs is similar for all samples when the orientation of the neighboring grains is the same (Fig. 6(a)). In contrast, the ordered SRO density in the ACTRs of low-T complexions and high-T complexions is different when the neighboring grains are oriented in different ways (Fig. 6(b)). To provide a single measure of the SRO asymmetry, we calculate the difference in ordered SRO density between the two ACTRs in Figs. 6(c) and 6(d). The ordered SRO density difference is calculated by subtracting the ordered SRO density of ACTR-1 from that of ACTR-2 for a given grain boundary. This means that a positive value of ordered SRO density difference means that ACTR-2 has a higher density of ordered SRO (and negative values mean ACTR-1 has a higher density of ordered SRO). The difference in ordered SRO density at the ACTRs is negligible compared to the absolute values for the symmetric boundaries (Fig. 6(c)). For the asymmetric low-T and high-T complexions, the difference in ordered SRO density is significant, with both positive and negative values measured (Fig. 6(d)).

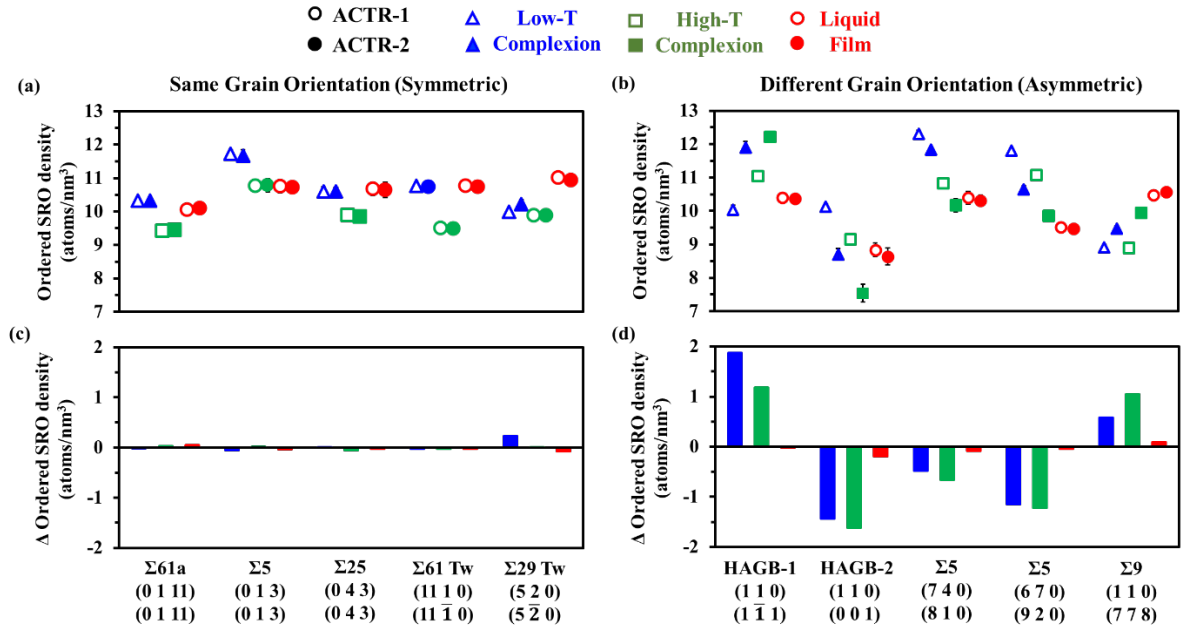

**Fig. 6.** Ordered SRO density in the ACTRs in different low-T complexion, high-T complexion, and wetting film models with the (a) same (symmetric) and (b) different (asymmetric) grain orientations. The difference between the SRO density of the ACTRs in different low-T complexions, high-T complexions, and liquid films with the (c) same (symmetric) and (d) different (asymmetric) neighboring grains. The structural order of the ACTRs in both low-T and high-T complexions is symmetric when the neighboring grains have the same grain orientation and asymmetric when the neighboring grains have different grain orientations. However, the structural order in the ACTRs of wetting films is symmetric even when the neighboring grains are different.

A closer inspection of Fig. 6 shows that the ordered SRO density in an ACTR is influenced not only by the crystalline grain that touches it. For example, the ordered SRO densities in ACTR-1 of HAGB-2 and Σ9 asymmetric grain boundaries are different, even though they have the same orientations for Grain 1, with a (110) plane terminating at ACTR-1. Furthermore, ACTR-1 has a higher ordered SRO density than ACTR-2 in HAGB-2 whereas the opposite trend is observed for Σ9 asymmetric grain boundaries. The only difference between these two asymmetric boundaries is Grain 2, highlighting that it is not only the nearby grain but rather some competition between the two confining crystals, where their differences manifest as different ACTR structures.

Asymmetric complexion structures have been observed for asymmetric boundaries [35, 84, 111], demonstrating the importance of local crystallography on the resultant complexion structures in terms of the neighboring grain. Our initial attempts at understanding SRO in amorphous complexions used a similar thought process where asymmetry should be dependent on the crystal structure/orientation next to the interface [85, 112]. However, our data definitely shows that the distribution of SRO within the ACTRs is dependent on both of the confining crystals. To understand how the difference between Grain 1 and Grain 2 impacts complexion structure, the mismatch or incompatibility between the crystals must be quantified. The planar density of neighboring grains along the interface is one of the parameters that has been shown to influence the structure of solid-liquid interfaces, where intermixing between the solid and liquid increased with a decrease in the packing density of the nearby solid [113]. However, planar density was not found to be a useful descriptor here. For example, the neighboring grains in the two Σ5 asymmetric grain boundary complexions have the same planar density but different values of ordered SRO density in the ACTRs. In a recent paper, Nathaniel et al. [114] used geometric phase analysis of transmission electron microscopy images to measure lattice strain next to the grain boundaries, which these authors used as a parameter to determine the defect absorption capacities of different interfaces. In another work, Lee et al. [115] showed that lattice strains arising from the neighboring grain misfit produce a defect nearby the grain boundary in order to relieve the accumulated distortional strain energy, increasing interface energy and restricting grain boundary migration in pure Cu.

The von Mises shear strain captures the amount of distortion for individual sites near the initial grain boundary structure, measured with respect to the ideal lattice, which is needed to accommodate the mismatch or incompatibility between the neighboring crystals. As shown above, the SRO of ACTRs is dominated by the ordered motifs, that are moderately distorted versions of the perfect FCC structure. Therefore, we hypothesize that the local distortional strains near the original grain boundary can be used to predict the resultant density of the distorted ordered SRO motifs once the amorphous complexion forms. The lattice strains near the initial grain boundary structure were calculated in the pure Cu bicrystals, with Figs. 7(a) and 7(b) showing examples from the Σ5 symmetric tilt grain boundary and HAGB-2, respectively. The atoms in the bicrystals are colored according to the von Mises shear strain, with our focus being the FCC atoms near the grain boundary since these can be ascribed to either Grain 1 or Grain 2. The nearby atoms on either side of the grain boundary have higher von Mises shear strain than the crystal interior due to mismatch between the neighboring grains. The high strain sites can be interpreted as a sign that the region in the grain near the boundary must accommodate the crystal incompatibility. The grain atoms in the vicinity of the grain boundary are likely to show more clear signs of grain incompatibility, so we restrict ourselves to the first and second nearest neighbors of the grain boundary in each of the confining crystals, shown in the insets outlined in green. A visual inspection can provide a sanity check, as the von Mises shear strain color pattern is symmetric around the Σ5 symmetric tilt grain boundary yet asymmetric for the HAGB-2 with different neighboring grains.

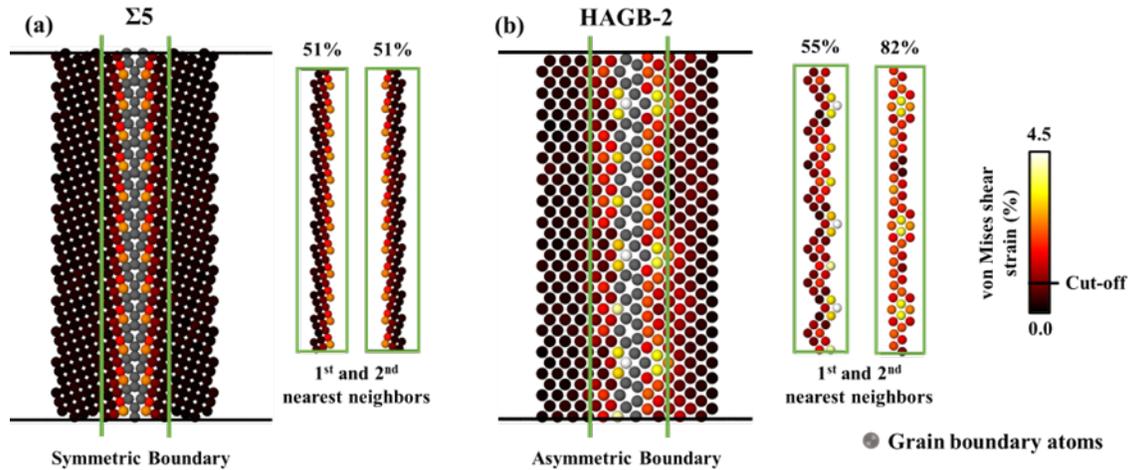

Fig. 7. The initial structure of (a) a Σ5 symmetric tilt grain boundary and (b) an asymmetric HAGB-2 grain boundary. The atoms are colored according to their atomic von Mises shear strain (grain boundary atoms appear gray) to highlight local distortions next to the grain boundary. Isolated views of the crystalline atoms which are the first and second nearest neighbors to the grain boundary atoms show significant variations in the atomic von Mises shear strain as compared to the crystal interior. The variations in this distortional strain are symmetric around the Σ5 symmetric tilt grain boundary and asymmetric for the asymmetric HAGB-2. A cutoff value of 1.1% was chosen to define highly distorted atoms, although the predictions of ordered SRO density that follow were not found to be sensitive to this specific choice.

The structural misfit in each grain was quantified by calculating the percentage of first and second nearest neighbors of grain boundary atoms with greater distortional strain, termed "highly distorted atoms", than the cut-off strain value of 1.1%, with the calculated percentages shown above the insets in Fig. 7. Figs. 8(a) and (b) show the percentage of highly distorted atoms for the various symmetric and asymmetric initial grain boundary configurations studied here, respectively. The percentage of highly distorted atoms on either side of the grain boundary is the same when the neighboring grains are the same (Fig. 8(a)). In asymmetric boundaries, the percentage of highly distorted atoms is clearly different on either side of the grain boundary (Fig.

8(b)). Thus, this type of lattice distortion analysis seems to capture the asymmetry that mimics the trends in SRO density.

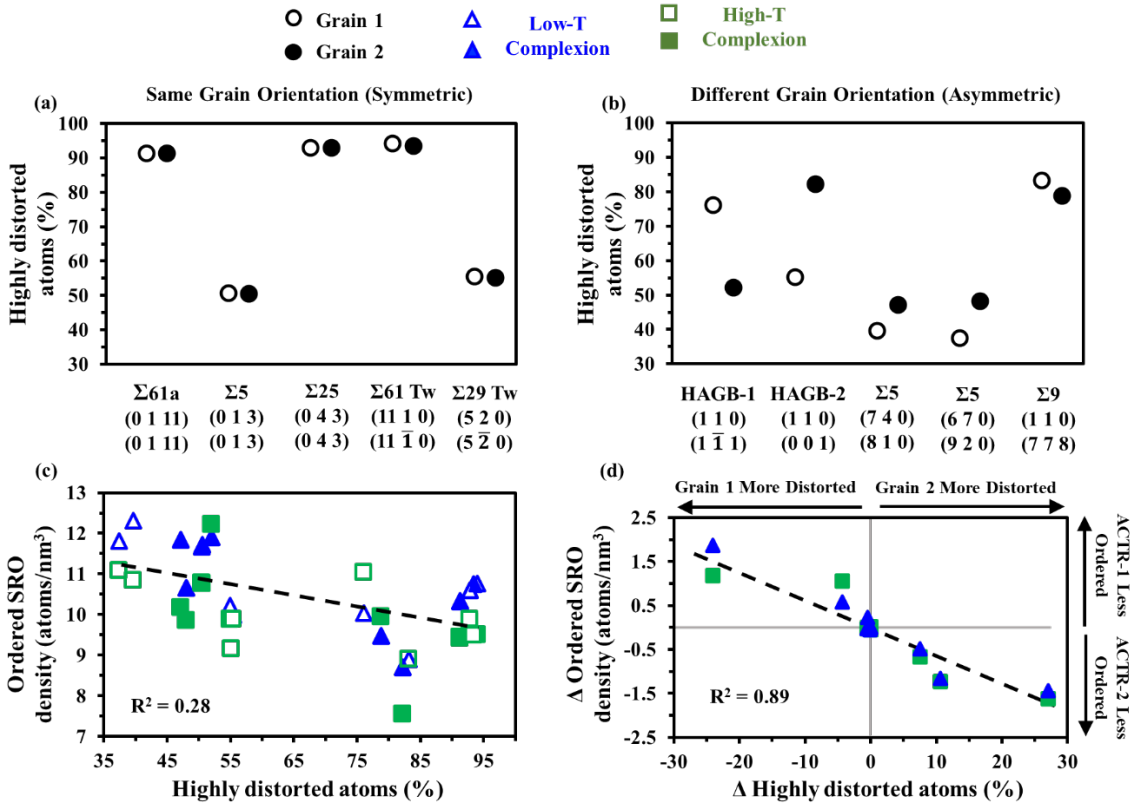

Fig. 8. Percentage of highly distorted atoms within the 1st and 2nd nearest neighbor shells of the grain boundary for bicrystals with (a) the same neighboring grains (symmetric boundaries) and (b) different neighboring grains (asymmetric boundaries). (c) Ordered SRO density in the complexion ACTRs (final state) plotted against the percentage of highly distorted atoms near the initial grain boundary (initial state), with the latter being a measure of grain incompatibility. (d) The difference in ordered SRO density of the complexion ACTRs (final state) plotted against the difference in the percentage of highly distorted atoms near the boundary in the original bicrystal (initial state). An inversely proportional relationship is observed in (d), demonstrating that grain incompatibility determines the distribution of ordered SRO between the ACTRs.

We first plot the ordered SRO density in the ACTRs versus the percentage of highly distorted atoms in the corresponding grain before doping in Fig. 8(c). The ordered SRO density in ACTR-1 or ACTR-2 is plotted with respect to the percentage of highly distorted atoms in either Grain 1 or Grain 2, respectively. The two parameters are not strongly correlated, again showing that the SRO of an ACTR cannot be described only by the nearest grain. To capture the incompatibility between two grains in a single parameter, we take the difference in the percentage of highly distorted atoms along the interface in Grain 2 and Grain 1 (similar to the discussion of the difference in ordered SRO density, a convention of Grain 2 minus Grain 1 is used). A positive value, therefore, means that Grain 2 is more distorted to accommodate the incompatibility between the two grains, and vice versa. The difference in ordered SRO density of the two ACTRs can then be plotted as a function of the difference in the percentage of highly distorted atoms between the two grains, as shown in Fig. 8(d). The difference in ordered SRO density and the percentage of highly distorted atoms are both near zero for all the symmetric grain boundaries, showing that there is no SRO asymmetry when the grains evenly distribute the local strain associated with the grain boundary. The difference in ordered SRO density of the ACTRs is clearly observed to be inversely proportional to the difference in the percentage of highly distorted atoms, with the data showing a high coefficient of determination value for the fitted linear curve ($R^2 > 0.89$). Although differences in ordered SRO density are observed between the low-T and high-T complexions, the neighboring grain incompatibility is found to dominate the SRO development here, as both low and high temperature complexion data fall along the trendline (the two states have the same grain incompatibility since the initial bicrystal is the same). The inverse relation between the two quantities means that the more disordered or strained crystal in the pair will have less ordered SRO density within the ACTR that develops. On the other hand, the grain which expressed less incompatibility with its neighbor will have a higher density of ordered SRO in the associated ACTR. As a whole, these findings clearly show that the incompatibility between grains determines the relative structural SRO of the amorphous complexions that nucleate in between them. The entire grain-film-grain system must be considered to understand the resultant SRO distribution of the complexion.

**3.3 Extension to a polycrystalline grain boundary network**

Bicrystals with isolated and flat grain boundaries are necessarily high-symmetry configurations as compared to the diversity of grain boundaries found in real materials. Therefore, to strengthen our hypothesis that grain incompatibility determines ACTR SRO on a broader scale, we perform a similar analysis on a Cu-Zr polycrystalline model with a random collection of grain boundaries. The variations in lattice strains near the grain boundaries are first examined in a pure Cu polycrystalline sample equilibrated at 900 K (initial state), following Ref. [79]. Fig. 9(a) shows a pure Cu polycrystal sliced along a (111) plane of the simulation cell where the grain atoms are colored according to the von Mises shear strain and the grain boundary atoms appear gray. Variations in the lattice strains of grain atoms adjacent to the grain boundary again highlight the incompatibility between the neighboring grains. Based on our hypothesis, this incompatibility should be closely related to the SRO of the ACTRs that nucleate in the polycrystalline Cu-Zr alloy. Fig. 9(b) shows a Cu-Zr polycrystal (final state) equilibrated through hybrid MC/MD simulation at 900 K, sliced along the same (111) simulation cell plane, with the atoms colored according to their local crystal structure. FCC atoms appear green and non-crystalline atoms appear white, while the atoms with ordered SRO type are black. Similar to our bicrystal observations, the ordered SRO is primarily observed along the ACTRs, with very few examples being observed in the complexion interiors.

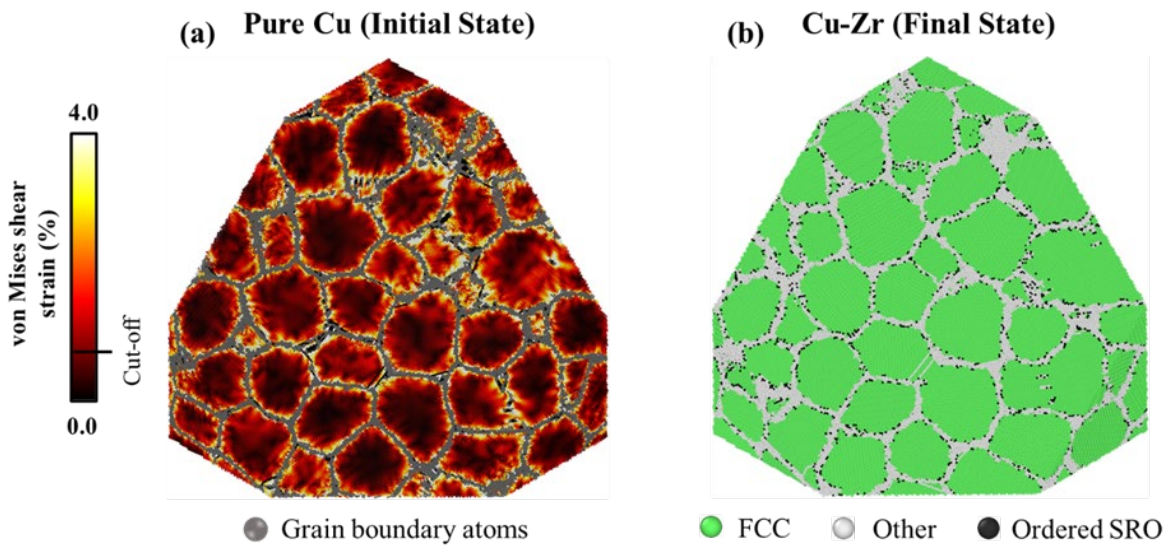

**Fig. 9. (a) Variations in the local von Mises shear strain for a pure Cu polycrystal model. The grain boundary atoms are colored gray to allow for visualization of variations in the atomic distortional strain of grain atoms near the interface. (b) Ordered SRO polyhedra, denoted by black atoms, are observed primarily at the ACTRs in a Cu-Zr polycrystal. In this image, FCC atoms are colored green and non-crystalline atoms are colored white.**

Fig. 10(a) shows a representative example of a general grain boundary in the pure Cu polycrystal (initial state) which transformed into an amorphous complexion after Zr segregation (Fig. 10(c)). Fig. 10(b) shows only the first and second nearest neighbor atoms of a general grain boundary (all the grain boundary and crystal interior atoms have been removed) to highlight the region being analyzed. Like the isolated bicrystal samples, the boundaries of ACTRs in a polycrystal can be defined at the location where the ordered polyhedra types become more common than the perfect FCC and disordered polyhedra types. In the example shown, the lattice strain and the number of ordered SRO motifs are asymmetric across the grain boundary, with Grain 1 being more distorted in the pure Cu configuration and then ACTR-1 having less ordered SRO in the Cu-Zr alloy after complexion nucleation.

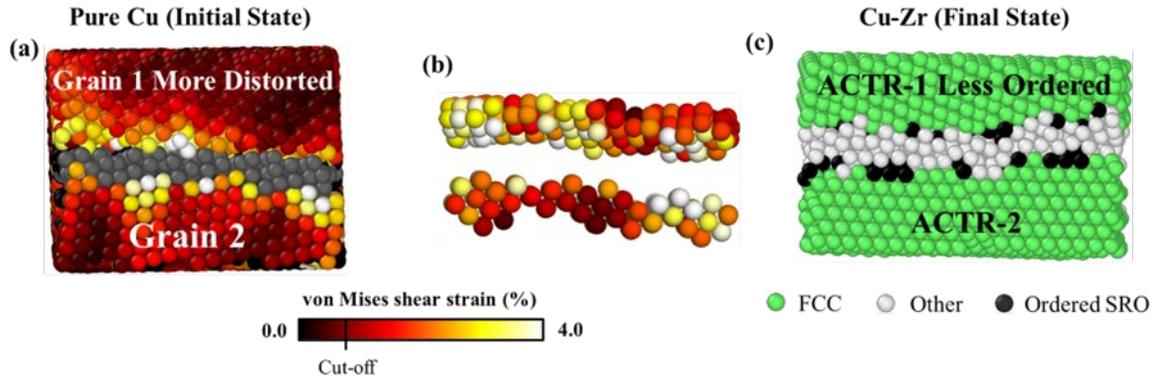

Fig. 10. A specific example of (a) a grain boundary in pure Cu which transformed into (c) an amorphous complexion in Cu-Zr polycrystal after applying MC/MD to add dopant elements and find a local equilibrium state. The distribution of distortional strain and ordered SRO polyhedra are asymmetric across the grain boundary and in the resulting amorphous complexion.

In the Cu-Zr polycrystal equilibrated at 900 K (final state), ~6% of the grain boundaries transformed into amorphous complexions upon solute segregation and are used in our analysis presented in Fig. 11. Both the grain incompatibility and the difference in ordered SRO density are calculated in the same way as described above for the bicrystal samples, to ascribe a directionality to the values. Due to the non-flat shape of the grain boundaries in the polycrystal, ACTR volume is measured as the sum of the Voronoi volume of all the atoms in the ACTR. An inversely proportional relationship is again found between the difference in ordered SRO density and the grain incompatibility. Thus, our understanding of complexion structure from the bicrystals can be successfully extended to the more general polycrystal case.

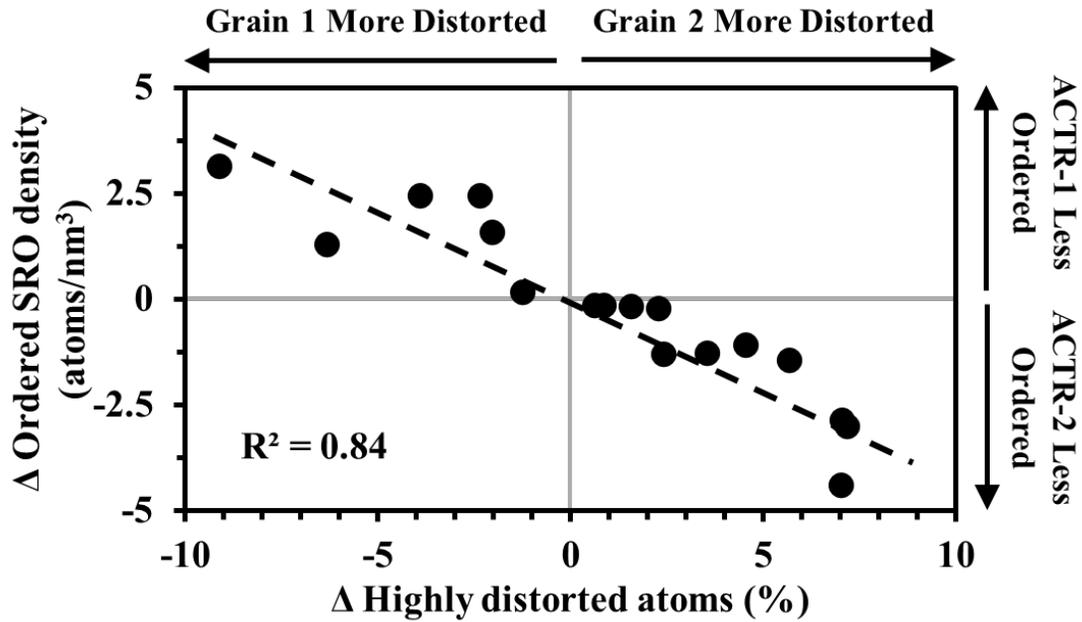

Fig. 11. The difference in ordered SRO density of the complexion ACTRs in the Cu-Zr model (final state) plotted against the difference in the percentage of highly distorted atoms in the pure Cu model (initial state). Again, the inverse relation between the two parameters shows that the structural order of the ACTRs can be predicted by the original incompatibility between the neighboring grains, consistent with the trend observed in the bicrystal configurations.

### 3.4 Complexion theory and future opportunities

The current understanding of interfacial complexions can be thought of as building on traditional grain boundary theory and models, yet important differentiating aspects exist. Grain boundaries have been commonly analyzed using different equilibrium thermodynamics-based models such as those of Fowler–Guggenheim [116], McLean [117], and Lee and Aaronson [118], which are solution-based models that treat grain boundaries as separate thermodynamic entities. Other models like diffuse-interface [119, 120] and force-balance [121, 122] models have had some success in explaining the transition of boundaries to wetting films [123, 124]. However, the common feature that these models treat the grain boundary region as a distinct phase is flawed for describing complexions and their structural transitions. Our work here unequivocally proves that amorphous complexions are not simply disordered second phases that happen to nucleate and get stuck at a grain boundary, but instead features whose structure and composition (and resultant properties) are intimately dependent on the confining grains. A liquid film that solidifies at a grain boundary would be the same as the wetting films from this study. Their structure has some variations through the thickness, as the interior and ACTR SRO are different, but does not recreate the symmetry/asymmetry of the crystallography associated with the confining grains. In contrast, the amorphous complexions studied in our work bear distinct signatures of the competition between the confining grains. This highlights that the entire grain-film-grain system must be considered, and emphasizes that complexions cannot be understood without considering the confining crystals as well as the incompatibility between these crystals.

Descriptors such as thickness are commonly used to label complexion types [25, 27, 125, 126], but this metric does not necessarily provide a pathway for easily distinguishing amorphous complexions (one complexion) from wetting films (a second phase bound by two complexions). While it is tempting to target a critical thickness as the boundary between these two complexion types, this feature alone is not a robust definition. As shown here and also reinforced in recent experimental studies from our research group [127, 128], amorphous complexions can be as thick as 8-10 nm, which is much larger than previously thought [25, 124]. The true differentiating factor is that an amorphous complexion and its confining grains find a local equilibrium, with prior studies showing that non-uniform film thickness along a single grain boundary can serve as a sign that a wetting film might exist [126]. Our work provides an even more robust type of proof by highlighting that spatial trends in local structural order enable the identification of complexion type. To accomplish this however, one needs to be able to measure SRO with high spatial resolution. While this is relatively straighforward when using atomistic simulations, this information has traditionally been extremely challenging to obtain from experiments. However, recent developments in materials characterization offer great promise on this topic. X-ray Bragg ptychography provides a method that can image small crystalline displacements and defect-induced strains with high spatial resolution [129-131]. In addition, new scanning transmission electron microscopy (STEM) methods show immense promise. For example, cepstral STEM, a new form of 4D-STEM, was used by Shao et al. [132] to detect fluctuations associated with lattice disorder and image the corresponding distortions caused by lattice mismatch in high entropy alloys. Recent work by Nathaniel et al. [114] measured lattice strains using high-resolution transmission electron microscopy combined with image processing techniques to determine incompatibilities between neighboring grains and understand the defect absorption capacity of the associated grain boundaries. Some of these new techniques have shown their promise by measuring local SRO in bulk metallic glasses. For example, Hirata and Chen [133, 134] characterized the local structure of metallic glasses using angstrom-beam electron diffraction and detected the local ordering among atomic first nearest neighbors. These authors used experimentally obtained diffraction patterns that were directly compared with the simulated diffraction patterns of atomic clusters from MD simulations to identify the structural motifs present in the amorphous materials. Pekin et al. [100] used 4D-STEM techniques to collect nanobeam electron diffraction information from metallic glass specimens during an in situ deformation experiment, showing that local structural order evolved near a stress-concentration prior to sample failure. As a whole, the integration of these recently developed experimental techniques with computational approaches should allow the field to provide more robust identification of complexion type as well as in depth studies of internal complexion structure in advanced engineering alloys.

## 4. Summary and Conclusions

In this work, the SRO of amorphous complexions was examined to test the hypothesis that the confining grains determine internal complexion structure. The structural order of the ACTR regions was found to be controlled by the incompatibility between the neighboring grains in both bicrystal and polycrystal models of Cu-Zr. The following conclusions can be drawn from this study:

- Ordered polyhedra dominate the SRO of ACTRs whereas disordered polyhedra are the most frequent in the complexion interior, demonstrating that the ACTRs exhibit clearer signatures of the order of the confining grains. The local structure of amorphous complexions was found to be independent of the cooling rates used in this study, although we note that relatively fast cooling rates must be used in atomistic simulations.
- The ordered SRO density between the two ACTRs is asymmetric when the confining crystals are different and symmetric when the neighboring grains are the same. The local structure of ACTRs is therefore representative of any mismatch that exists between the confining crystals. However, any SRO asymmetry vanishes in the wetting films.
- The incompatibility between the confining crystals is quantified by calculating the percentage of highly distorted atoms within the first and second nearest neighbors of grain boundary in each grain. The difference in ordered SRO density of the ACTRs is inversely proportional to the incompatibility between the neighboring grains, showing that the structure of ACTRs is dependent on both grains that surround the complexion, not just the closest grain.

Overall, this work provides a more complete understanding of the internal structure of amorphous complexions and the role of the confining crystals on these features. Our findings demonstrate that amorphous complexions are structurally different from wetting films and highlights a robust method of distinguishing these two films from one another. As a whole, our findings emphasize that the grain-film-grain system must be treated as a single unit to formulate a meaningful understanding of a grain boundary complexion.

**Acknowledgments**

This research was supported by the U.S. Department of Energy, Office of Science, Basic Energy Sciences, under Award No. DE-SC0021224. Structural analysis and atomic visualization were performed with software supported by the National Science Foundation Materials Research Science and Engineering Center program through the UC Irvine Center for Complex and Active Materials (DMR-2011967).

[39] G.J. Tucker, D.L. McDowell, Non-equilibrium grain boundary structure and inelastic deformation using atomistic simulations, International Journal of Plasticity 27(6) (2011) 841-857.
[40] A. Hasnaoui, H. Van Swygenhoven, P. Derlet, Cooperative processes during plastic deformation in nanocrystalline fcc metals: A molecular dynamics simulation, Physical Review B 66(18) (2002) 184112.
[41] Y. Wang, J. Li, A.V. Hamza, T.W. Barbee, Ductile crystalline–amorphous nanolaminates, Proceedings of the National Academy of Sciences 104(27) (2007) 11155-11160.
[42] Y.M. Wang, A.V. Hamza, T.W.B. Jr., Incipient plasticity in metallic glass modulated nanolaminates, Applied Physics Letters 91(6) (2007) 061924.
[43] C.M. Grigorian, T.J. Rupert, Thick amorphous complexion formation and extreme thermal stability in ternary nanocrystalline Cu-Zr-Hf alloys, Acta Materialia 179 (2019) 172-182.
[44] A. Khalajhedayati, Z. Pan, T.J. Rupert, Manipulating the interfacial structure of nanomaterials to achieve a unique combination of strength and ductility, Nature Communications 7(1) (2016) 1-8.
[45] J.L. Wardini, C.M. Grigorian, T.J. Rupert, Amorphous complexions alter the tensile failure of nanocrystalline Cu-Zr alloys, Materialia  (2021) 101134.
[46] Z. Pan, T.J. Rupert, Amorphous intergranular films as toughening structural features, Acta Materialia 89 (2015) 205-214.
[47] S. Pal, K. Vijay Reddy, C. Deng, On the role of Cu-Zr amorphous intergranular films on crack growth retardation in nanocrystalline Cu during monotonic and cyclic loading conditions, Computational Materials Science 169 (2019) 109122.
[48] J. Bernal, Geometry of the structure of monatomic liquids, Nature 185(4706) (1960) 68-70.
[49] J. Bernal, J. Mason, Packing of spheres: co-ordination of randomly packed spheres, Nature 188(4754) (1960) 910-911.
[50] P.H. Gaskell, A new structural model for amorphous transition metal silicides, borides, phosphides and carbides, Journal of Non-Crystalline Solids 32(1) (1979) 207-224.
[51] D.B. Miracle, W.S. Sanders, O.N. Senkov, The influence of efficient atomic packing on the constitution of metallic glasses, Philosophical Magazine 83(20) (2003) 2409-2428.
[52] D.B. Miracle, The efficient cluster packing model – An atomic structural model for metallic glasses, Acta Materialia 54(16) (2006) 4317-4336.
[53] Y. Zhang, F. Zhang, C. Wang, M. Mendelev, M. Kramer, K. Ho, Cooling rates dependence of medium-range order development in C u 64. 5 Z r 35. 5 metallic glass, Physical Review B 91(6) (2015) 064105.
[54] J.J. Maldonis, A.D. Banadaki, S. Patala, P.M. Voyles, Short-range order structure motifs learned from an atomistic model of a Zr50Cu45Al5 metallic glass, Acta Materialia 175 (2019) 35-45.
[55] P. Ramachandrarao, On glass formation in metal-metal systems, International Journal of Materials Research 71(3) (1980) 172-177.
[56] T.D. Shen, U. Harms, R.B. Schwarz, Correlation between the volume change during crystallization and the thermal stability of supercooled liquids, Applied Physics Letters 83(22) (2003) 4512-4514.
[57] Y. Cheng, E. Ma, Atomic-level structure and structure–property relationship in metallic glasses, Progress in Materials Science 56(4) (2011) 379-473.
37